\documentclass[11pt]{article}
\usepackage{amsmath}
\usepackage{amssymb}
\usepackage[dvips]{epsfig}

\def\##1{\underline #1}
\def\=#1{\underline{\underline #1}}

\def\.{\mbox{ \tiny{$^\bullet$} }}

\begin{document}

%\newpage
%\setcounter{page}{1}

\noindent

 \vskip 0.4cm

\begin{center}
{\large {\bf Comment on `Towards gravitationally assisted negative
refraction of light by vacuum' \cite{LM}}}

{\em Martin W. McCall\footnote{Fax: +44 (0)20 7594 7714;
m.mccall@imperial.ac.uk}}

{\em Department of Physcs, The Blackett Laboratory\\
Imperial College London\\
Prince Consort Road, London SW7 2BZ, UK}

%September 20th, 2004
\end{center}

\begin{center}
{\bf Abstract}
\end{center}

Reference \cite{LM} claims that matter--free regions of spacetime
may support electromagnetic waves for which the phase velocity
3-vector is oppositely directed to the power flow 3-vector. This
was the definition of negative refraction adopted in \cite{ML}. We
here show that this cannot occur in regions of spacetime for which
the electromagnetic energy density is positive.
%Media capable of supporting such waves are often
%referred to variously as `Left--handed media' \cite{Schultz1,
%Schultz2, Smith}, `backward media' \cite{Lindell},
%`double--negative media', \cite{Ziol}.

\section{Covariant formulation of `Negative Refraction'}
In complex notation, the stress energy tensor associated with the
electromagnetic field is given by
\begin{equation}
\label{stresstensor} T_{\alpha\beta} =
\frac{1}{8\pi}\left[F_\alpha^{~\gamma}\left(F_{\beta\gamma}\right)^*
-\frac{1}{4}g_{\alpha\beta}F_{\mu\nu}\left(F^{\mu\nu}\right)^*\right]~,
\end{equation}
where taking the real part is understood, and the units are
Gaussian with $c=1$. The electromagnetic field tensor may be
written in terms of the vector potential via
\begin{equation}
\label{vectorpotential}
F_{\alpha\beta} = \nabla_{\alpha}A_\beta -
\nabla_{\beta}A_\alpha~.
\end{equation}
Maxwell's equations are
\begin{equation}
\label{maxwell1}
\nabla_\beta F^{\alpha \beta} = 4\pi J^\alpha~,
\end{equation}
and
\begin{equation}
\label{maxwell2} \nabla_{[\gamma}F_{\alpha\beta]} = 0~.
\end{equation}
In terms of the vector potential (\ref{maxwell2}) is satisfied
automatically, whilst under the Lorentz gauge ($\nabla_\alpha
A^\alpha =0$) equation (\ref{maxwell1}) becomes
\begin{equation}\label{Maxwellvector1}
\nabla^\alpha\nabla_\alpha A^\beta - R_\delta^{~\beta}A^\delta =
-4\pi J^\beta~,
\end{equation}
where $R_\delta^{~\beta}$ is the Ricci tensor. In vacuum,
neglecting any curvature induced by the field, both
$R_\delta^{~\beta}$ and $J^\beta$ vanish leaving
\begin{equation}\label{covariantmaxwell}
\nabla^\alpha\nabla_\alpha A^\beta =0~.
\end{equation}
A plane wave solution of Maxwell's equations may be written as
\begin{equation}
\label{planewave} A^\alpha = C^\alpha e^{iK_\mu X^\mu}~,
\end{equation}
where $C^\alpha$ is a covariantly constant 4-vector
amplitude\footnote{In flat spacetimes global solutions to
Maxwell's equations exist
 for which the amplitude $C^\alpha$ is covariantly constant over all spacetime. In curved spacetime it is
 often possible to
 construct solutions in which $C^\alpha$ varies slowly - the so--called `geometric optics
approximation' (\cite{wald}, p.71; \cite{mtw}, p.570).}, and
$K_\mu$ are the covariant components of the 4-wavevector,
 $\left(\omega ,{\bf k}\right)$. We take $\omega >0$.

Negative refraction is said to occur whenever $k_j$ is oppositely
signed to $T_{0j}$.

\section{Flat spacetime}
Consider a region of spacetime where, with respect to a given
coordinate system, the metric coefficients are approximated by
uniform functions. Can such a spacetime region support `negative
refraction' as defined above?
%The metric of spacetime varies from one event to another but is
 %covariantly constant. To say that the metric coefficients are constant over a region of spacetime
 %(as in \cite{LM})
 %only makes sense when referred to a particular coordinate system.  Suppose for example globally flat
 %vacuum is described by cartesian
 %spatial coordinates and a global galilean time. The metric
 %is everywhere Minkowskian, $g_{\alpha\beta} = \eta_{\alpha\beta}$. Now insert
 %a rotating mass $M$ somewhere. At a sufficient distance $r$ from the mass the (time--averaged)
 %metric in the same cartesian frame
 %is now given by\footnote{Need to check the units} \cite{plebanski}
%\begin{equation}
%g_{00} = 1 - \frac{2k}{c^2}\frac{M}{r}~,~~ g_{0j} =
%-\frac{2k}{c^3}\frac{1}{r^3}\epsilon_{jkl}X^kL^l~, ~~g_{jk} =
%-\delta_{jk}\left(1-\frac{2k}{c^2}\frac{M}{r}\right)~,
%\end{equation}
%where $L^k$ are the components of the mass's angular momentum
%3-vector. Provided the spatial extent of the region under
%consideration is not too large, then the $g_{\alpha\beta}$ may be
%approximated by constant functions in the given frame. In this
%region $g_{\alpha\beta ,\gamma}$ (and all higher derivatives)
%vanish under this approximation. We are thus considering an
%extended region of {\em flat} spacetime in which metric
%coefficients do not vary with cartesian coordinates (or with
%time).
%Such a flat region is
%conformally equivalent to one in which the metric is given by
%$\eta_{\alpha\beta}$.
Since in the considered region all covariant derivatives may be
replaced by ordinary derivatives, we have from equations
(\ref{stresstensor}), (\ref{covariantmaxwell}), (\ref{planewave})
and the Lorentz gauge ($A^\mu_{~,\mu}=0$) that
\begin{eqnarray}
F_{\mu\nu} &=& i\left( K_{\mu}A_\nu -
K_{\nu}A_\mu\right)~,\label{Fmunu}\\
K_{\mu}K^{\mu} &=& 0~,\label{dispersion}\\
K^{\mu}A_{\mu} &=& 0~.\label{eigenmode}
\end{eqnarray}
Using equations (\ref{stresstensor}), (\ref{Fmunu}),
(\ref{dispersion}) and (\ref{eigenmode}) it is straightforward to
show that in this case
\begin{equation}
T_{\alpha\beta} = \frac{1}{8\pi} A^\mu
\left(A_\mu\right)^*K_{\alpha}K_{\beta}~,
\end{equation}
and hence that
\begin{equation}
T_{0j} = \frac{1}{8\pi}A^\mu \left(A_\mu\right)^*\omega k_j~.
\end{equation}
Now since $A^\mu \left(A_\mu\right)^*
>0$ (on account of $T_{00}>0$), then $k_j$ must take the same sign as $T_{0j}$.
The negative refraction condition cannot be fulfilled in regions
of flat spacetime.

\section{Curved spacetime}
Within the geometric optics approximation, the above argument
works with covariant derivatives replacing ordinary derivatives,
and is thus valid for curved spacetime also. We conclude that
regions of spacetime in which the electromagnetic energy density
is positive cannot support `negative refraction'.

\end{document}